\documentstyle{article}
\begin{document}
\large
\baselineskip=0.8cm
\input psfig
\title{Grafted Rods: A Tilting Phase Transition}
\author{F. Schmid $^{\ddag}$, 
  D. Johannsmann$^{\dag}$ and A. Halperin $^{*}$ \\
\hspace*{3cm}\\
$^{\ddag}$ Institut f\"ur Physik, Johannes-Gutenberg Universit\"at, 
55099 Mainz, Germany \\
$^{*}$ ICSI, 15 rue Jean Starcky, BP 2488, 68057 Mulhouse, France \\
$^{\dag}$ Max-Planck-Institut f\"ur Polymerforschung, Postfach 3148, 55021 Mainz, Germany
}

\date{}
\maketitle

{\bf Abstract} -
A tilting phase transition is predicted for systems comprising rod 
like molecules which are irreversibly grafted to a flat surface, so 
that the non interacting rods are perpendicularly oriented.
The transition is controlled by the grafting density $\rho$.
It occurs as $\rho$ increases as a result of the interplay between two 
energies. Tilt is favoured by the van-der-Waals attraction between the rods.
It is opposed by the bending elasticity of the grafting functionality.
The role of temperature is discussed, and the tilting mechanism
is compared to other tilting transitions reported in the literature.

\vspace{2cm}

\noindent
PACS-Numbers: 64.70, 68.35, 82.65

\newpage

\section{Introduction}

Anisotropic molecules forming thermotropic and lyotropic liquid crystals
can exhibit tilt. The onset of tilt is the characteristic of the 
smectic A-smectic C phase transition\cite{LC}. Tilted phases are observed 
in monolayers formed by amphiphilic molecules\cite{Amp}.
A variety of tilting phase transitions are possible at solid-fluid 
interfaces incorporating anisotropic molecules\cite{TLC}-\cite{li}.
Diverse molecular mechanisms can give rise to such phase transitions.
In the following, we propose a novel mechanism for the development of tilt at 
surfaces carrying grafted rods. In particular, we consider a flat solid 
surface supporting covalently bound rodlike molecules at a surface density 
$\rho < 1$. 
The rods are grafted to the surface so that isolated rods are 
oriented along the surface normal. The grafting functionality, the ``joint",
is assumed to allow bending at the price of an elastic energy penalty.
A tilting phase transition is predicted to occur as the grafting density 
$\rho$ increases. The tilt is favoured by the van-der-Waals attraction 
between the rods, but is opposed by the bending penalty.
The interplay between these two energies gives rise to a tilting phase 
transition. The transition is predicted at zero temperature. However,
the mechanism still works at sufficiently low finite temperatures,
as we shall argue in the discussion.

From a fundamental perspective, this mechanism is of interest as a complement 
to the variety of tilting mechanisms studied in the literature. 
Tilting transitions are also of importance vor the design of alignment 
layers for liquid crystal displays\cite{LCD}. The orientation of the nematic 
director inside a liquid crystal cell with no electric or magnetic field 
applied is usually determined by the interface. Properly designed interfaces 
supply the desired in-plane asymmetry as well as a certain "pretilt" angle. 
Grafted liquid crystal (LC) layers may form good 
alignment layers, when the LC-director adopts the tilt of the grafted layer. 
In practice, the tilt of the nematic director may not be equal to the tilt 
of the grafted layer due to the complicated interaction between the 
nematic medium and the layer. This does not however detract from the
practical potential of this system. A tilting phase transition 
driven by the grafting density may provide a scheme to fabricate 
alignment layers with an arbitrary tilt angle.

Another aspect of grafted layers of interest to
liquid crystal alignment is their behavior under nematic 
orientational stress. Such stresses occur during the Frederiks transition and 
are of high practical importance as well. Coupling between the tilting phase 
transition and the Frederiks transition is expected to reduce the switching
voltage of such devices\cite{TLC}.

There are a number of reports, where rods have been covalently attached to 
a surface \cite{bue,sek,kno}. In most cases, however, the joint is assumed 
to be so flexible that it will not show any resistance to bending. Recently, 
attachment with rigid joints has been claimed as well\cite{whi,mac}. Note 
however, that a rigid joint does not necessarily imply perpendicular 
orientation. The angle of the joint will depend on details of the chemistry. 
Right-angled joints are the exception rather than the rule. If, however, 
the tilt angle is small and the in-plane direction of tilt is random, 
it is reasonable to describe the rods as perpendicularly oriented.

The essential physics involved is discussed in section II for a highly 
simplified case: A linear array of grafted rods. An analysis of the two 
dimensional case is presented in section III. A comparison to previously 
discussed tilting phase transitions is given in the discussion.

\section{The Linear Array}

The underlying physics of the tilting transition are most easily understood 
in the case of a linear array. The rods, of length $L$ and diameter $a$, are 
grafted onto a straight line at regular intervals of width $D$. In the 
absence of rod-rod interactions, the grafting functionality imposes 
perpendicular orientation of the rods, {\em i.e.} the angle between the rods 
and the surface normal is $\gamma =0$. However, the joint can be deformed 
with a bending elastic penalty of
\begin{equation}
\label{II1}
U_{bend} =\sum_{i=1}^{\infty} A_{i}{\sin}^{2i} \gamma,
\end{equation}
where $A_{i}$ are non negative phenomenological constants specified in units 
of energy. The rods also experience mutual van-der-Waals attraction. We 
picture the rods as comprising monomers of size $a$\cite{JNI}. The non 
retarded van-der-Waals attraction between two monomers separated by a 
distance $r$ is
\begin{equation}
w=-c/r^6,
\end{equation}
where $c$ is a material constant with units of ${\rm [energy][length]^6}$. 
The van-der-Waals attraction between two parallel rods, $E(D,\gamma)$, may 
be obtained by integration of the pairwise interactions between monomers 
belonging to the different rods (see Appendix A). At $D/L < 1$, the leading 
terms in $E(D,\gamma)$ are 
\begin{equation}
\label{II3}
 E(D, \gamma) = - \frac{K}{(d \cos \gamma)^5} \{
\frac{3 \pi}{2} + d \cos \gamma (\cos^2 \gamma - 3 - 3 \gamma \tan \gamma)
+ {\cal O}(d^5) \}
\end{equation}
with $d=D/L$ and $K = c/(4 a^2 L^4)$. 
Note that $E(D,\gamma)$ favors large tilts since in this limit the reduced 
distance between the rods is $D \cos \gamma < D$. Notice further that 
$E(D, \gamma)$ diverges at $\gamma = \pi/2$, when the rods 
are fully tilted. This last feature is however an artefact, since 
$\gamma = \pi/2$ is actually unattainable because of excluded volume 
interactions between the rods. An upper bound for the maximal possible tilt 
in the $D/L < 1$ case is ${\gamma}_{up} = \arccos (a/D)$. 

To analyze the phase behaviour, it is convenient to consider $E(D, \gamma)$ 
in the limit of $d \ll 1$ and $\gamma \ll 1$. In this limit, where 
corrections due to the finite length of the rods are negligible, the 
leading terms in the expansion of of $E(D, \gamma)$ in powers of 
${\sin \gamma}^2$ are
\begin{equation}
\label{II4}
E(D, \gamma) = -(3 \pi K/2d^5)[1 + (5/2){\sin}^2 \gamma + 
(35/8){\sin}^4\gamma+ \cdots] =-\sum_{i=1}^{\infty}B_i {\sin}^{2i}\gamma.
\end{equation}
We now consider the total van-der-Waals energy, $U_{vdW}$, 
of a rod in a uniformly tilted array. To obtain $U_{vdW}$ we sum the pairwise 
interactions of the rod {\em i.e.} 
$U_{vdW}=2\sum_{n=1}^{\infty} E(nD, \gamma)$. In the limit of $D/L \ll 1$, 
when all terms in $E(D, \gamma)$ exhibit a $D^{-5}$ dependence, the result is
\begin{equation}
\label{vdw}
U_{vdW} = 2 \zeta(5) E(D, \gamma),
\end{equation}
where $\zeta(5) = \sum _{i=1}^{\infty} k^{-5} \approx 1.037$ is the 
appropriate Riemann zeta function. In doing so, we assume that the van der 
Waals interactions are additive. That is the case when the dielectric 
susceptibility of the materials is an additive function of the constituting 
components {\em i.e.} when the susceptibility is proportional to the 
concentration of the rods\cite{Add}. One should however note that this 
requirement in not always fulfilled\cite{NAd}. If one considers as an
approximation only nearest neighbour interactions, the calculated
interaction energy $U_{vdW} = 2 E(D,\gamma)$ is very close to 
(\ref{vdw}). Hence the total van der Waals energy is dominated by the
nearest neighbour contributions.

The behaviour of the linear array of rods is determined by the total energy 
per rod $U=U_{bend}+U_{vdW}$. It is instructive to consider the behaviour of 
$U$ in a number of scenarios. If one overlooks the excluded volume effect 
between the rods, $U_{vdW}$ diverges at $\gamma =\pi/2$. When  $U_{bend}$ 
is finite at $\gamma = \pi/2$, this divergence dominates the behaviour of $U$,
$U(\gamma=\pi/2)=-\infty$, indicating that all the rods are tilted.
While $U(\gamma)$ may exhibit two minima, $U(\pi/2)=-\infty$ always 
corresponds to the equilibrium state in this scenario. A richer repertoire
is possible when the bending penalty at $\gamma=\pi/2$
is infinite as well. In that case, all terms in the expansion of $U_{bend}$
have to be retained. Consider, for example, the situation when the 
coefficients of ${\sin}^{2i}\gamma$ of $U_{bend}$ and $U_{vdW}$ are such that
$A_i-B_i>0$ for all $i \ge 2$. A second order tilting phase
transition is expected when $A_1-B_1=0$ {\em i.e.} $A_1/K=15\pi/4d^5$ with 
$d=D/L$ or $d=d_c=(15\pi K/4A_1)^{1/5}$. The rods are perpendicular to the 
surface, $\gamma=0$, when $d>d_c$, while for $d<d_c$ they are tilted at 
angle of $\gamma \propto \sqrt{d_c-d}$. A first order phase transition is 
expected at $A_1-B_1>0$, if some of the higher order terms are
negative and large enough such that 
$\sum_{i \ge 2} (A_i-B_i) \sin^{2i}\gamma <0$
for some value of $\gamma < \gamma_{up}$.

The role of higher order terms of $U_{bend}$ is certainly important.
A discussion based on this effect suffers however from two disadvantages.
Typically, the higher order $A_i$'s are unknown. Also, a chemical design
allowing for control of these parameters is impractical.
To simplify the analysis, we assume that the bending energy is Hookean
up to a cutoff angle $\gamma_{up}$. For specifity, we assume that it is
the result of excluded volume interactions and hence given by
$\gamma_{up} = \arccos (a/D)$ as noted earlier. Note however $\gamma_{up}$
may also be imposed by the bending potential.

It is helpful to consider first $U$ in the limit of $d \ll 1$ and 
$\gamma \ll 1$, retaining only the first order terms of $U_{vdW}$ and 
$U_{bend}$ as specified by (\ref{II1}) and (\ref{II4}). $U$ is then given by
\begin{equation}
U=[A_1-15\pi K\zeta(5)/2d^5]{\sin}^2 \gamma.
\end{equation}
When $\epsilon =A_1-15\pi K\zeta(5)/2d^5>0$, the system may exhibit a first 
order tilting phase transition. We shall discuss this scenario in greater 
detail shortly. A second order transition is expected when $\epsilon=0$,
while if $\epsilon < 0$ the system will be tilted. To analyze the first 
order transition, it is helpful to express $U$ in terms of (\ref{II1}) and 
(\ref{II3}). The energy per rod in the limit of $d \ll 1$ is then 
\begin{equation}
U = A_1 {\sin}^2\gamma - \alpha/(d^5 {\cos}^5 \gamma),
\end{equation} 
where $\alpha = 3 \pi \zeta(5) K$ and $\gamma$ is allowed to vary 
between $0$ and${\gamma}_{up}$. $U$ is minimal at $\gamma =0$, where
$U(0)=-\alpha/d^5$, and at ${\gamma}_{up}$, where 
$U(\gamma_{up})=A_1{\sin}^2 {\gamma}_{up}-\alpha/(d^5{\cos}^5{\gamma}_{up})$.
It is maximal at $0<{\gamma}_{max}<{\gamma}_{up}$, defined by
${\cos}^7{\gamma}_{max} =5\alpha/2A_1 d^5$.
%
%
A first order phase transition between an untilted phase and a phase tilted 
at an angle ${\gamma}_{up}$ occurs when $U(0) = U({\gamma}_{up}) $ or
$\alpha/d^5 A_1 = {\sin}^2{\gamma}_{up}/({\cos}^{-5}{\gamma}_{up}-1)$.

\section{The Planar Case}

To extend our considerations to two dimensional, planar arrays, it is
necessary to allow for two geometrical effects. First, for a given rod the
number of interacting neighbors increases with the distance. Thus, a shell 
of radius $r$ and thickness $dr$ contains $\rho r dr$ interacting rods, while 
in the linear array the number is always $\rho dr$. Second, in the two 
dimensional case the shortest distance between parallel rods, $\delta$, 
varies with azimuthal angle, $\phi$, between the vector joining the grafting 
sites and the direction of the tilt. When $\phi=0$ the shortest distance 
is $\delta=r\cos \gamma$ while for $\phi = \pi/2$ it is $\delta=r$. In general, 
$\delta$ is given by $\delta = r(1-{\cos}^2\phi {\sin}^2 \gamma)^{1/2}$ 
(Figure 1). In realistic situations, it is also necessary to allow for the 
randomness of the grafting sites. These are unlikely to form a regular 
lattice. It is thus useful to characterize the array of grafting sites by a 
pair distribution function
\begin{equation}
g(r) = \rho h(r).
\end{equation}
The introduction of $g(r)$ also simplifies the mathematical treatment of 
the problem. The van-der-Waals interaction energy per rod in a uniformly 
tilted array is
\begin{equation}
\label{av}
U_{vdW}=\rho \int_0^{\infty} rdr \int_0^{2 \pi}d \phi h(r) V(r,\phi,\gamma),
\end{equation} where $V(r, \phi,\gamma)=E(r, \theta)$ and $\theta$ is 
defined by $\sin \theta = \sin \gamma \cos \phi$. It is then possible to 
specify $U_{vdW}$ in terms of the inverse moments of $h(r)$
\begin{equation}
\label{utot1}
{\mu}_3 = \int_0^L h(r)dr/r^3 \qquad \qquad {\mu}_4 = \int_0^L h(r)dr/r^4,
\end{equation}
leading to 
\begin{equation}
\label{utot2}
U_{vdW}(\rho,\gamma) =
-\frac{\rho c \pi}{ a^2} \mu_3 
[{}_2F_1(\frac{1}{2},\frac{5}{2},1,\sin^2 \gamma) \frac{\pi}{2} l 
- \frac{1}{\cos^4 \gamma}] + {\cal R}(\gamma)
\end{equation}
with the hypergeometric function 
${}_2F_1(1/2,5/2,\sin^2\gamma) = 
\int_0^{\pi/2} d \phi [1-\sin^2 \gamma \cos^2 \phi]^{-5/2}$ 
and $l=L (3\mu_4)/(2 \mu_3)$.
The remainder ${\cal R}$ is of order $1/L^2$ (see Appendix B).

For simplicity, we take $h(r)$ to be a simple step function
\begin{equation}
h(r) = \Theta(r - D)
\end{equation}
with the minimal distance between rods $D$. At $D \ll L$, the inverse moments
$\mu_i$ are then given by $\mu_i = 1/((i-1)D^{i-1})$ and $l = L/D$.
We further assume that the minimal distance between rods depends on
the grafting density according to a proportionality law
$\rho = \nu/D^2$. This implies that the rod grafting sites are distributed 
evenly on the plane, like in the case of the linear array discussed in 
section 2, where the distance between rods is kept fixed.
The van-der-Waals energy can then again be expanded
in powers of $\sin \gamma$.
\begin{equation}
\label{vdwexp}
U_{vdW}(\rho,\gamma) = 
- \tilde{K} L \sqrt{\rho}^5 \{1 + (5/4)\sin^2 \gamma + (105/64) \sin^4 \gamma 
+ \cdots \} = - \sum_i \tilde{B}_i \sin^{2i} \gamma
\end{equation}
with $\tilde{K} = c \pi/(4 a^2 \sqrt{\nu}^3)$. 
After combining this expression with the bending energy per rod 
(\ref{II1}), one can follow the discussion from the previous 
section: A second order transition is expected at $A_1 = \tilde{B_1}$, 
{\em i.e.} at the critical density
\begin{equation}
\rho = \rho_c = (\frac{4 A_1}{5 \tilde{K}})^{2/5} \cdot L^{-2/5}.
\end{equation}
It may be preempted at some lower density by a first order tilt transition,
if the cutoff angle $\gamma_{up}$ is large enough such that
\mbox{$U_{vdW}(\rho_c,\gamma_{up})-U_{vdW}(\rho_c,0)+U_{bend}(\gamma_{up})=0$}.
The van der Waals energy in the limit $D/L \ll 1$ scales
like $L \sqrt{\rho}^5$, whereas the bending energy per rod is independent
of rod length and grafting density. Hence the densities, at which
phase transitions occur, scale with the rod length $L$ like
$\rho_c \propto L^{-2/5}$.

We note that these considerations only hold under the
supposition that the rods are distributed evenly on the plane,
$D \propto 1/\sqrt{\rho}$, as mentioned above. If the 
rods are grafted independently of each other, a Poisson distribution
is expected. In this case, the assumption $D=a$ with the monomer size $a$
may be more appropriate. Eqn (\ref{vdwexp}) then
has to be replaced by
\begin{equation}
U_{vdW}(\rho,\gamma) = 
- \frac{c \pi}{4 a^5} \rho L 
\{1 + (5/4)\sin^2 \gamma + (105/64) \sin^4 \gamma + \cdots \} 
\end{equation}
and densities of phase transitions scale with the rod length as 
$\rho_c \propto 1/L$. A second order phase transition is expected at 
$\rho_c = (16 a^5 A_1)/(5 \pi c L)$. However, in this case, the validity
of a mean field averaging as performed in eqn (\ref{av}) becomes 
questionable. This is because domains with different tilt order and 
local defects of the tilt order parameter may become important even at 
zero temperature. 

\section{Discussion}

The tilting behaviour considered in this paper results from the interplay
between three factors: The van-der-Waals attraction between the rods, the
bending elasticity of the joint and the grafting constraints. Clearly no 
joint is perfectly rigid. The necessary grafting constraints involve two 
ingredients:
(i) The bonding of the rods' ends to a flat surface in a manner imposing
perpendicular orientation of the non interacting rods.
(ii) The absence of lateral mobility.  
The rods start to tilt at the minimal grafting density $\rho_c$, where the 
effect of the attractive interaction starts to overcome the bending penalty.
Upon further increasing the grafting density, the excluded volume 
interactions between rods gain importance. As the rods are pushed together, 
they gradually stand up and finally undergo another continuous transition 
to an untilted state at $a=D$, where $\gamma_{up} = \arccos (a/D) = 0$. 
This second transition is driven by the interplay between attractive and 
repulsive interaction. Thus one expects tilt order in a well-defined density 
regime (Figure 2).

We should note that, at zero temperature, there is no entropic opposition 
to the in plane symmetry breaking associated with collective tilt. At
finite temperatures, however, the entropy comes into play, giving
rise to a variety of new phenomena:

First, it is well known that Goldstone excitations will destroy any
true long range order in two dimensional systems with continuous
symmetry, where the ground state is continuously degenerated\cite{mw}.
In our case, the Goldstone modes are long wavelength fluctuations
of the in plane tilt direction.
At sufficiently low temperatures, however, one still has quasi long
range order, {\em i.e.} the correlation functions decay algebraically.
Unless preempted by a first order transition, Kosterlitz-Thouless-type
unbinding of vortex excitations is expected to destroy this order at high
temperatures. Note that the energy contained in the vortices will 
approach zero as the tilt goes to zero. Therefore systems with a small 
tilt will undergo a Kosterlitz-Thouless transition at lower temperatures than
systems with high tilt. This implies that the tilting phase behaviour at 
finite, sufficiently low, temperatures is very similar to the zero 
temperature behaviour -- but with two differences: (i) the tilted state now 
has quasi long range order rather than true long range order, and (ii)
the second order tilting transitions are replaced by Kosterlitz Thouless 
transitions.

Second, the entropy being another factor which opposes collective tilt, 
it may assume the role of the bending elasticity at
finite temperatures. Indeed, the competition of attractive interactions 
between rods and rod orientational entropy gives rise to a scenario of 
tilting transitions which is very similar to the one discussed 
here\cite{wang,tilt}. The details may be different because the entropy 
will depend on the lateral density due to the packing constraints, whereas the 
bending elasticity is independent of lateral density.

When lateral mobility is allowed, phase separation will replace the tilting 
phase transition (cf. \cite{li,l1}), because the surface
free energy of untilted rods is higher than that of tilted rods.
Note however that this last scenario can be modified by several factors. 
First, if the rods are given some flexibility, the loss of surface free energy 
per chain upon lowering the chain density is compensated in part by 
the gain of conformational entropy. As a consequence, phase 
separation occurs at lower surface densities. Self consistent field 
calculations show that, for sufficiently flexible chains, the areal density 
of the condensed phase may be low enough to favour tilt\cite{l2}. 
Second, the phase separated condensed phase may support tilt even in systems
of perfectly stiff rods if this is energetically rewarded, e.g. by attractive 
chain-surface interactions\cite{wang,chen,Ahalperin,rice}. 
Third, the distance of closest approach between grafting sites may be larger 
than $a$, the rod diameter. This last situation is reminiscent of the onset of 
tilt in amphiphilic monolayers, when the distance between ``grafting sites" is 
determined by the effective radius of the ionic head group\cite{shih}.
A similar mechanism \cite{RC} for the development of tilt has been proposed 
for lamellae formed by rod-coil diblock copolymers {\em i.e.} polymers 
consisting of a rod like block joined to a flexible, coil like block. These 
can form lamellae in a selective solvent which is a precipitant for the rods.
In this case the surface free energy of the close packed rods opposes the 
tilt which is favoured by the repulsive interactions between the overlapping 
coils. These last tilt mechanisms are however less closely related to our 
system since they do not involve rods grafted to a surface with $\rho <1$. 
The rods are always closely packed. Furthermore, the driving force for the 
tilt comes from units outside the rods (substrate, head groups and flexible 
blocks respectively) and will not operate in a system consisting solely of 
rods.

As discussed in the introduction, tilting transitions are particularly
important, from a technical point of view, in relation to liquid crystal 
alignment. If the surface
is in contact with a nematic liquid crystal, the tilt of the rods at
the surface will influence the director in the bulk of the liquid crystal, 
and vice versa.  Tilt transitions such as the one discussed here will then 
translate into ``anchoring transitions'', which involve changes of the
direction of alignment in the bulk of the liquid crystal.

In this context, it is instructive to briefly discuss the tilting phase 
transition predicted for liquid crystalline polymers (LCPs) grafted 
onto a flat surface immersed in a good nematic solvent\cite{TLC}.
In this case the tilt results from the interplay between the repulsive 
monomer-monomer interactions and the elasticity of the solvent.
This situation involves a number of distinct features. 
First, the effect is expected in the case of long, main chain LCPs supporting 
a number of hairpin defects. Such LCPs adopt a highly anisotropic, ellipsoidal 
form such that their major axis is aligned with the nematic director.
Second, the effect depends on the anchoring conditions imposed by the
grafting surface. It is predicted when the anchoring is homogeneous {\em i.e.}
the nematic director at the interface is parallel to the surface. As a result 
of the coupling between the LCPs and the nematic solvent, the orientation of 
the isolated LCPs is with their major axis parallel to the surface.
As the grafting density increases, the overlap between the LCPs grows.
This, in turn, results in an increase of the repulsive interactions between 
the LCPs. These can be weakened by tilting the LCPs thus decreasing the 
monomer volume fraction within the layer. The onset of tilt is however 
opposed by the elasticity of the nematic solvent since it results in a 
distortion of the nematic medium. The interplay of these two contributions 
gives rise to a second order tilting phase transition. Tilt is
now favoured by repulsive interactions rather than by attraction. The 
role of the joint is assumed by the nematic solvent. The coupling between 
the LCP and the oriented nematic determines the orientation of the isolated 
LCPs and gives rise to the elastic penalty that opposes tilt.

As compared to the case studied in the paper and the tilting transitions 
discussed so far, the last scenario is different in that the tilting 
transition does not involve in-plane symmetry breaking. The azimuthal 
symmetry is already broken in the ``untilted'' state, in which the
LCPs lie parallel to the surface. This is an effect of the nematic
medium, which insures collective alignment even at the absence of
direct interaction between the molecules. Such tilting transitions 
without symmetry breaking are frequently found in liquid crystal systems.

An experimental example are anchoring transitions induced by
trans-cis isomerizations of azobenzene molecules, which are covalently attached 
to the surface\cite{bue,sek,kno} and immersed in a nematic LC solvent. In these 
cases, the rods are attached to the surface via flexible spacers comprising 
4-10 CH${}_2$-units. UV--irradiation induces a transition from a rod like 
trans conformation to a highly bent cis conformation in the azobenzene 
molecules. The rods in the trans state favour perpendicular (homeotropic)
alignment of the liquid crystal molecules, whereas the substrate and the 
molecules in the cis state favour parallel (homogeneous) alignment. At a
high density of trans units, the competition of these two factors leads to
tilted anchoring. UV--irradiation causes the density of trans units to
decrease, thereby inducing a continuous transition to an untilted state with 
parallel anchoring\cite{bue}.

\section{Conclusions}
We have described a tilting transition in grafted layers of rigid rods attached
to the surface via a chemical bond, which resists bending. The competition 
between bending elasticity on the one hand and van der Waals attraction on 
the other hand drives the transition. The transition can be of first or
second order. We find perpendicular alignment for low coverage, and tilted 
alignment for higher densities. Second order transitions will be preempted
by Kosterlitz-Thouless transitions at nonzero temperatures. We compare the 
transition with a number of related tilting transitions, which have been 
discussed in the context of monolayers and of liquid crystal alignment. 
In some cases similar scenarios are found, where the entropy takes the role 
of the stiff chemical bond.

\section*{Acknowledgements}

The authors greatfully acknowledge stimulating discussions with Michael B\"uchel

\section*{Appendix A: The one dimensional case}

We consider two parallel rods of length $L$ grafted at distance 
$D$ from each other, and uniformly tilted by an angle $\gamma$. The sum
of van der Waals interactions between monomers from
different rods is then given by
\begin{equation}
E(D,\gamma) = - (c/a^2) \int_0^L dx \int_0^L dx'
\frac{1}{[(D \cos \gamma)^2 + (x-x'+D \sin \gamma)^2]^3} .
\end{equation}
We can decompose the integral and rewrite $E(D,\gamma)$ as
\mbox{$E = - c/(4 a^2 L) \cdot \widehat{E}(D/L,\gamma)$} with
\begin{eqnarray*}
\lefteqn{\widehat{E}(d,\gamma) = 8[\int_0^1 dx \frac{1-x}{(\delta^2+x^2)^3}}\\
&&- \int_0^\varepsilon dx (\varepsilon-x) \Big[ \frac{1}{(\delta^2+x^2)^3} 
-\frac{1/2}{(\delta^2+(1+x)^2)^3}-\frac{1/2}{(\delta^2+(1-x)^2)^3} \Big],
\end{eqnarray*}
where $\delta = d \cos \gamma$ and $\varepsilon = d \sin \gamma$.
The integrals can be solved exactly, and one obtains
\begin{eqnarray}
\lefteqn{\widehat{E}(d,\gamma) = \frac{1}{(d \cos \gamma)^4} \Big\{
\cos^2 \gamma - 3 \gamma \tan \gamma - 
\frac{d^2 \cos^2 \gamma (1+d^2)}{(1+d^2)^2 - 4 d^2 \sin^2 \gamma}}\\
&&+ \frac{3}{2} \Big[
(\frac{1+d \sin \gamma}{d \cos \gamma}) 
\arctan (\frac{1+d \sin \gamma}{d \cos \gamma}) 
+ (\frac{1-d \sin \gamma}{d \cos \gamma}) 
\arctan (\frac{1-d \sin \gamma}{d \cos \gamma}) \Big] \Big\}.\nonumber
\end{eqnarray}
This function can be expanded in different ways for the cases
$d < 1$ and $d > 1$:

\noindent
(a) $d < 1$:
\begin{equation}
\label{e1}
\widehat{E}(d,\gamma) = \frac{1}{(d \cos \gamma)^4}
\Big\{\frac{3 \pi}{2 d \cos \gamma} + \cos^2 \gamma  
- 3 - 3 \gamma \tan \gamma \Big\}
+ {\cal R}_1(d,\gamma)
\end{equation}
with
${\cal R}_1(d,\gamma) = 2/5 + d^2(4 - 32/7 \cdot \cos^2 \gamma) 
+ {\cal O}(d^4)$

\noindent
(b) $d > 1$:
\begin{equation}
\widehat{E}(d,\gamma) = \frac{4}{d^6} + \frac{1}{d^8}(14-16 \cos^2 \gamma)
+ {\cal O}(\frac{1}{d^{10}}) \equiv {\cal R}_2(d,\gamma)
\end{equation}

\section*{Appendix B: The planar case}

More generally, let the two rods be grafted on the $xy$-plane and
uniformly tilted in the $y$-direction. The vector connecting their grafting
points can be parametrized as \mbox{$\vec{r}=r (\cos \phi, \sin \phi,0)$} 
and the tilt vector is $(\sin \gamma, 0,\cos \gamma)$. The interaction between 
the two rods is then $V(r,\phi; \gamma) = E(r,\theta)$ with an effective
tilt angle $\theta$ determined by $r \sin \theta = \vec{r} \vec{t}$
(see Figure 2).
We now consider an assembly of rods, which are randomly distributed on the
plane according to an isotropic pair distribution function 
$g(r) = \rho h(r)$. The total van der Waals energy per rod, 
given by eqn (\ref{utot1}), can then be calculated explicitly for
the leading terms in the expansion (\ref{e1}). Using the identities
\begin{displaymath}
\int_0^{\pi/2} \frac{d \phi}{\sqrt{1-z^2}^5} = \frac{\pi}{2}
{}_2F_1(\frac{1}{2},\frac{5}{2},1,\sin^4 \gamma),
\end{displaymath}
with $z = \sin \gamma \cos \phi$, and
\begin{displaymath}
\int_0^{\pi/2} d \phi \Big[
\frac{3 z \arcsin z}{\sqrt{1-z^2}^5} + \frac{z^2+2}{(1-z^2)^2} \Big]
= \frac{\pi}{\cos^4 \theta},
\end{displaymath}
one obtains eqn (\ref{utot2}). 
The order of magnitude of the neglected rest ${\cal R}$ can be estimated by
\begin{eqnarray*}
\frac{4 a^2 L^2}{\rho c} {\cal R} &=&
\int_0^{2 \pi} d \phi \: \{ \int _0^1 dx \: x \: h(x L) {\cal R}_1(x,\theta) +
\int _1^{\infty} dx \: x \:  h(x L) {\cal R}_2(x,\theta) \} \\
& \le &
2 \pi h_{max} \{ \int_0^1 dx \: x \: (2/5+4 x^2 + \cdots) 
+ \int_1^{\infty} dx \: x \: (4/x^6 + 14/x^8 + \cdots) \} \\
&=& {\cal O}(1),
\end{eqnarray*}
hence ${\cal R}$ is of order $1/L^2$.

\newpage

\section*{Figures}

\begin{figure}[h]
\unitlength=1mm
\caption{Two parallel rods grafted on the $xy$-plane}
\begin{picture}(140,110)
\put(0,0){
\psfig{figure=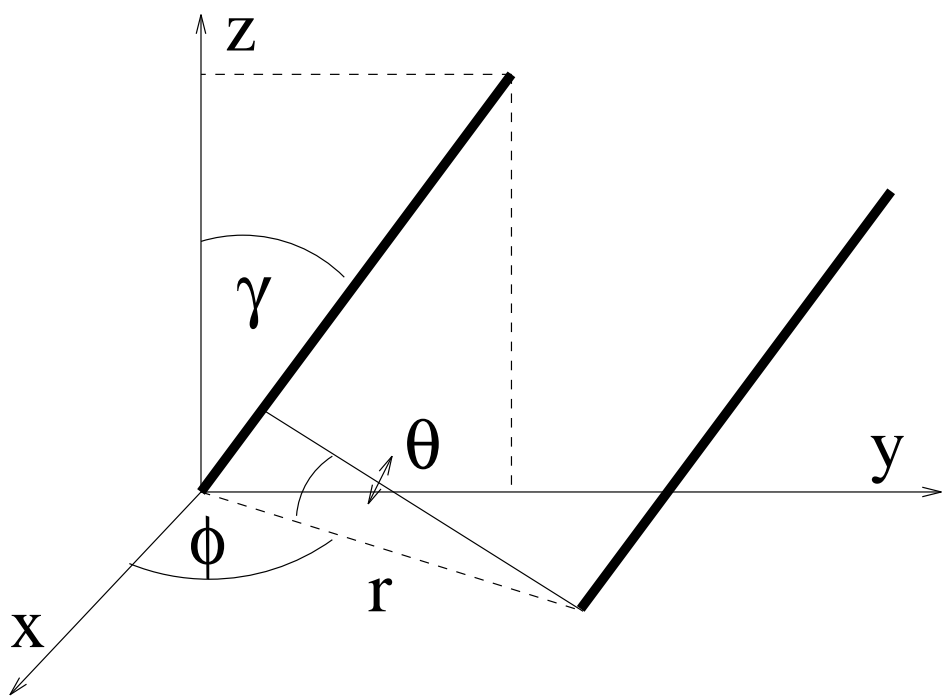}
}
\end{picture}
\end{figure}

\begin{figure}[h]
\unitlength=1mm
\caption{A schematic representation of the tilting transitions upon 
increasing grafting density $\rho$ at low temperatures}
\begin{picture}(140,110)
\put(0,0){
\psfig{figure=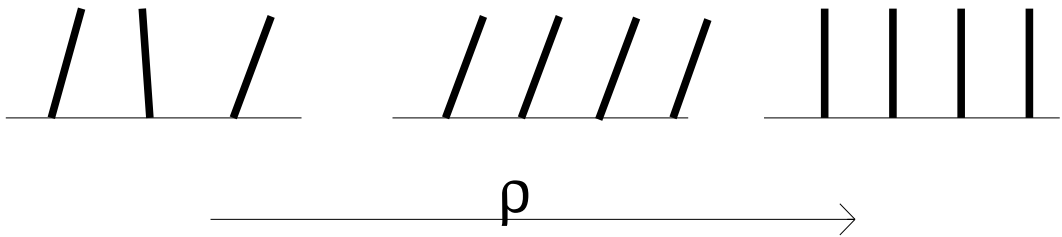,width=140mm,height=30mm}
}
\end{picture}
\end{figure}

\end{document}